\documentclass[sigconf]{acmart}
\usepackage{graphicx}

\usepackage{subcaption} %
\usepackage{makecell} %
\usepackage{xcolor} %
\usepackage{systeme} %
\usepackage{booktabs} %
\usepackage{siunitx}  %
\usepackage{amsmath}
\usepackage{array}

\makeatletter
  \newcommand\EatSpacesHack{\@bsphack\@esphack}
\makeatother
  \renewcommand{\comment}[1]{\EatSpacesHack}
  \newcommand{\PostSubmission}[1]{\EatSpacesHack}
  \newcommand{\todo}[1]{\EatSpacesHack}
  \newcommand{\basi}[1]{\EatSpacesHack}
  \newcommand{\zeyu}[1]{\EatSpacesHack}
  \newcommand{\ak}[1]{\EatSpacesHack}
  \newcommand{\reviewfix}[1]{\EatSpacesHack}

  \usepackage[suppress]{color-edits}
  \addauthor{ak}{blue}

\newcommand{\V}[1]{\mbox{\textit{#1}}}  %
\newcommand{\Vs}[1]{\mbox{\textit{\small #1}}}  %

\def\Snospace~{\S{}}

\author{Basileal Imana}
\email{imana@princeton.edu}
\orcid{0000-0002-6645-7850}
\affiliation{%
  \institution{Princeton University}
  \city{Princeton}
  \state{New Jersey}
  \country{USA}
}

\author{Zeyu Shen}
\email{zs7353@princeton.edu}
\orcid{0009-0000-2858-9214}
\affiliation{%
  \institution{Princeton University}
  \city{Princeton}
  \state{New Jersey}
  \country{USA}
}

\author{John Heidemann}
\email{johnh@isi.edu}
\orcid{0000-0002-1225-7562}
\affiliation{%
 \institution{USC/Information Sciences Institute}
 \city{Los Angeles}
 \state{California}
 \country{USA}
}

\author{Aleksandra Korolova}
\email{korolova@princeton.edu}
\orcid{0000-0001-8237-9058}
\affiliation{%
 \institution{Princeton University}
  \city{Princeton}
  \state{New Jersey}
  \country{USA}
}

\copyrightyear{2025}
\acmYear{2025}
\setcopyright{cc}
\setcctype{by}
\acmConference[FAccT '25]{The 2025 ACM Conference on Fairness, Accountability, and Transparency}{June 23--26, 2025}{Athens, Greece}
\acmBooktitle{The 2025 ACM Conference on Fairness, Accountability, and Transparency (FAccT '25), June 23--26, 2025, Athens, Greece}\acmDOI{10.1145/3715275.3732170}
\acmISBN{979-8-4007-1482-5/2025/06}

\title{External Evaluation of Discrimination Mitigation Efforts in Meta's Ad Delivery}

\begin{document}

\begin{abstract}
The 2022 settlement between Meta and the U.S. Department of Justice to resolve allegations of discriminatory advertising resulted is a first-of-its-kind change to Meta's ad delivery system aimed to address algorithmic discrimination in its housing ad delivery.
In this work,
  we explore direct and indirect effects
  of both the settlement's choice of terms
  and the Variance Reduction System (VRS) implemented by Meta on the actual reduction in discrimination.
  
We first show that the settlement terms allow for an implementation
  that does not meaningfully improve access to opportunities for individuals.
The settlement measures impact of ad delivery in terms of impressions, instead of unique individuals reached by an ad;
  it allows the platform to level down access,
  reducing disparities by decreasing the overall access to opportunities;
  and it allows the platform to selectively
  apply VRS to only small advertisers. 

We then conduct experiments to evaluate VRS with real-world ads,
  and show that while VRS does reduce variance,
  it also raises advertiser costs (measured per-individuals-reached),
  therefore decreasing user exposure to opportunity ads 
  for a given ad budget.
VRS thus \emph{passes the cost of decreasing variance to advertisers}. 

Finally,  
  we explore an alternative approach to achieve the settlement goals,
  that is significantly more intuitive and transparent than VRS.
We show our approach outperforms VRS by both increasing
  ad exposure for users from \emph{all} groups and reducing cost to advertisers, thus demonstrating that the increase in cost to advertisers when implementing the settlement is not inevitable.
  
Our methodologies use a black-box approach that relies on
  capabilities available to any regular advertiser, rather than on privileged access to data,
  allowing others to reproduce or extend our work.
\end{abstract}

\maketitle

\section{Introduction}

The 2022 agreement between the US Department of Justice (DoJ)
  and Meta to implement a Variance Reduction System (VRS)
  has been widely celebrated as a groundbreaking step
  in regulating ad delivery algorithms 
  in social media platforms to mitigate
  discrimination in domains of important life opportunities, such as housing and
  employment~\cite{FacebookvsHUD2, FacebookvsHUD2023}.
The need for changes in ad delivery algorithms due to their biases was demonstrated by
  investigative journalism~\cite{ProPublicaGender, ProPublicaRace},
  civil-rights audits~\cite{FbCivilRightsAudit2020, FacebookCivilAuditProgress},
  and academic research~\cite{Sweeney2013, Lambrecht2016, Ali2019a, Imana2021}.
The works highlighted the significant role that machine learning used in platforms' ad delivery systems
  can play in perpetuating discriminatory access to economic
  opportunities.
Although the settlement focused exclusively on housing ads,
  Meta has voluntarily expanded the deployment of VRS to 
  employment and credit ads~\cite{FacebookvsHUD2023}.

Any fairness intervention should consider two questions:
First, is the metric of algorithmic fairness effective at assessing a reduction in harm?
Second, what are the trade-offs for other objectives, such as utility~\cite{Hu2020, Baumann2024, Peysakhovich2023}?
The DoJ/Meta settlement suggests minimizing \emph{variance}
  between the demographic distribution of impressions in an
  ad's actual audience compared to the demographic distribution of impressions in the ad's \emph{eligible audience}.
For metrics, the settlement establishes
  a compliance target: 
  VRS should ensure that the variance
  of a specified proportion of ads
  does not exceed a certain threshold~\cite{vrscomplianceJune24}.
(We expand on settlement terms,
  Meta's implementation, and compliance requirements in \autoref{sec:background}).
However, both the chosen metrics and their implementation
  could result in trade-offs,
  with the ad reaching fewer recipients
  at a higher per-recipient cost,
  perhaps not achieving the best possible reduction in harm~\cite{Baumann2024}.

The first contribution of our work is to demonstrate that
  the settlement terms allow for an implementation that does not improve exposure to opportunity
  ads for individuals (\autoref{sec:design_critique}).
We first show that the settlement's goal is to balance ad impressions,
  not ad \emph{reach}, i.e. the actual number of unique recipients (\autoref{sec:imp_not_indi}).
Variance in impressions can be reduced by 
  repeatedly showing an ad to the same users,
  making the actual societal benefit unclear.
Second, the settlement specifies targets in terms of the proportion of ads for whom the variance should be limited, %
without taking into account their volume of 
  impressions or reach; thus ads with, say, 1k and 1M impressions
  count equally to the target. 
We show that 
  this metric allows for selective application of VRS to small advertisers (\autoref{sec:coverage_limitation}), thus potentially applying it to much fewer people than the goals may imply.
We quantify the potential effect of such selective application on access
  to opportunities by using public data from Meta on advertising budgets and reach.
Finally,
  we show the settlement requirements allows for a leveling down effect (\autoref{sec:critique_level_down}),
  where fairness is accomplished by reducing
  the outcomes of higher-performing groups
  down to the level of lower-performing groups,
  with no demographic group benefiting in the process~\cite{Mittelstadt2023}.

Our second contribution is to show that the specific VRS implementation chosen by Meta  
  makes ads more expensive for advertisers,
  therefore effectively reducing how many recipients see opportunities
  for a fixed ad budget (\autoref{sec:method1}).
We demonstrate this outcome through the first independent evaluation
  of VRS's impact on delivery of real-world ads.
We conduct our experiments using a novel black-box methodology that
  isolates VRS's role by running paired campaigns of the same ad with and without VRS applied.
Our findings show that while VRS reduces variance
  according to the legal compliance metrics for housing ads
  compared to a case without VRS intervention,
  fewer unique users are reached by the ad versions to whom VRS is applied and
  the cost of compliance is passed on to advertisers,
  making it more expensive for economic
  opportunity ads to reach a wider audience.

Our final contribution is to explore a new budget-splitting approach
  that outperforms Meta's VRS implementation by increasing exposure of opportunity ads 
  to \emph{all} demographic groups and reducing cost to advertisers relative to VRS (\autoref{sec:method2}).
Our findings show VRS's reduced utility 
  for users and advertisers is an artifact of
  Meta's implementation choices and not
  an inherent necessity
  for mitigating discrimination in ad delivery.
We explore budget-splitting as an example
  strategy that addresses the shortcomings of VRS's implementation
  while better meeting fairness and utility goals, and, unlike the current implementation,
  having the benefit of transparency and explainability.

We make data from all our experiments publicly available at~\cite{VRSAdDeliveryDataset}.


\section{Background}
\label{sec:background}

We summarize the terms of the Meta/DoJ settlement,
  the VRS implementation to fulfill these terms,
  and Meta's compliance reporting.
We base this overview on the public settlement terms~\cite{FacebookvsHUD2},
  Meta's white-paper and academic publication~\cite{FacebookvsHUD2023TR, Timmaraju_2023},
  and documents provided by the DoJ and an external third-party
  reviewer~\cite{FacebookvsHUDCase}.

\subsection{Settlement  Between DoJ and Meta}
\label{sec:settlement}

\paragraph{VRS to Reduce Variance in Ad Delivery}
A key settlement requirement is that Meta will implement a system to reduce variance
  by race and gender
  by aligning the demographics
  of an ad's \emph{actual audience} with that of an ad's \emph{eligible audience}.
This baseline is central to VRS's guarantees;
  we define it here and
  analyze its implications for fairness in ad delivery
  in \autoref{sec:design_critique}.
 
The  \emph{eligible audience} is the set of all users who fit the targeting criteria chosen
  by the advertiser and have received one or more impressions of any type of ad
  on Meta during the last thirty days~\cite{FacebookvsHUD2}. 
The baseline against which the VRS system measures variance is defined via the \emph{eligible ratio}, which relies on the eligible audience.
Specifically, the eligible ratio for a specific demographic group $g$ for a particular ad is calculated using the proportion of impressions
  received by users from the ad's eligible audience who belong to $g$ compared to impressions received by users from the ad's entire eligible audience from any advertiser on Meta in the last 30 days.
Mathematically, for a specific $ad$ and demographic group $g$,
  the eligible ratio is given by:
$\textbf{Eligible Ratio}_{g, ad} ={(\sum_{u \in \Vs{ad\_eligible\_aud} \cap g} \textbf{Imps}_{u})} / {(\sum_{u \in \Vs{ad\_eligible\_aud}} \textbf{Imps}_{u})}$,
where $\textbf{Imps}_{u}$ is the number of impressions that the user $u$ received from all advertisers on Meta over the last thirty days~\cite{FacebookvsHUD2023TR, Timmaraju_2023}. 

After an ad starts running, VRS measures the demographic distribution
  of the \emph{actual audience} of the ad, which is the set of all users in the eligible audience to
 whom at least one impression of this ad is displayed.
VRS then calculates a \emph{delivery ratio} for each demographic group $g$,
  which is the fraction of total impressions for an ad that were shown
  to members of $g$. Mathematically,
  $\textbf{Delivery Ratio}_{g, ad} = \frac{\textbf{Imps}_{g, ad}}{\textbf{Imps}_{ad}},$
where $\textbf{Imps}_{g, ad}$ is the number of impressions of the ad delivered to users in group $g$, and
  $\textbf{Imps}_{ad}$ is the total number of impressions for the ad.

Using the Eligible and Delivery Ratios,
  VRS then aims to reduce the \emph{variance} between the eligible and actual audiences of an ad, aiming for this variance to be below a 10\% or 5\% threshold.
The variance for an $ad$ is defined separately for gender and race:

\small
$ 
 \textbf{Variance}(\V{Gender})_{\Vs{ad}} = \\ \hspace{1em} \frac{1}{2}\sum_{g \in \{\text{Male, Female}\}} |\textbf{Eligible Ratio}_{g, ad} - \textbf{Delivery Ratio}_{g, ad}|
$

\vspace{1em}
$
 \textbf{Variance}(\V{Race})_{\Vs{ad}} = \\  \hspace{1em} \frac{1}{2}\sum_{\substack{g \in \{\text{African American, Hispanic,} \\ \text{White, Other}\}}} |\textbf{Eligible Ratio}_{g, ad} - \textbf{Delivery Ratio}_{g, ad}|,
$
\normalsize
where ``Other'' refers to all races other than African American, Hispanic and White~\cite{FacebookvsHUD2023TR, Timmaraju_2023}.
If an ad impression is delivered to a user with ``unknown'' gender,
  the impression is omitted in the calculation of variance for gender~\cite{vrscomplianceJune24}.

The variance metric, roughly-speaking, represents the minimum fraction of impressions that need to be moved
  between the groups for the delivery ratio to match the eligible ratio.
For example, if the delivery ratio is (0.4, 0.6) for males and females, respectively, and the eligible ratio is (0.5, 0.5),
 then the variance is $\frac{1}{2}(|0.5 - 0.4| + |0.5 - 0.6|) = 0.1$,
  indicating VRS needs to move a $0.1$ fraction of impressions from females to males
  to match the eligible ratio.
  
Gender for the computations is based on user self-report. Meta infers user race using Bayesian Improved Surname Geocoding (BISG), a method that gives probabilistic estimates for race based on surname and zip code~\cite{metatech, Elliott2009}.
We discuss open questions about how the use of BISG may impact VRS's performance in \autoref{sec:bisg}.

\paragraph{Compliance Metrics and Coverage}
The extent to which the variances need to be bounded is specified in    
  agreed upon \emph{compliance metrics}~\cite{vrscomplianceJune24}.
\emph{Coverage} is a key metric,
  defined as
  the percentage of housing ads among all housing ads run over the compliance reporting period of 4 months whose delivery variance falls below
  the thresholds of 5\% or 10\%.
Coverage targets are defined separately for gender and race and for ads that receive more than 300 and more than 1,000 impressions. The precise targets are given in~\autoref{tab:coverage}.
For example, for housing ads with 1,000+ impressions,
  VRS must ensure that the variance by gender is below 10\% for 91.7\% of ads and the variance by race is below 10\% for 81\% of the ads.

\begin{table}[h]
\centering
\begin{tabular}{lcc}
 & \multicolumn{2}{c}{\textbf{Coverage for ads that received:}} \\
\cmidrule(lr){2-3}
\textbf{Variance} & \(\geq 300\) Impr. & \(\geq 1,000\) Impr. \\
\midrule
Gender ($\leq10\%$) & 90.2\% & 91.7\% \\
Gender ($\leq5\%$) & 78.3\% & 84.5\% \\
Estimated race ($\leq10\%$) & 80.1\% & 81.0\% \\
Estimated race ($\leq5\%$) & 56.8\% & 61.0\% \\
\end{tabular}
\caption{Coverage Requirements for Housing Ads~\cite{vrscomplianceJune24}.}
\vspace{-3.1em} %
\label{tab:coverage}
\end{table}

\subsection{Meta's Implementation of VRS}
  \label{sec:background_meta_implem}

VRS is invoked for any ad the advertiser
  self-identifies as housing, employment, or credit (HEC)~\cite{FacebookvsHUD2023TR, FacebookvsHUD2023}.
Meta also separately examines whether ads may be on the HEC topics and refuses to run them without
  HEC tagging.
Once an ad tagged as HEC launches, VRS starts periodically measuring variance and
   \emph{adjusting the bids} for users on the advertiser's behalf, so as to increase
   delivery rate to a group currently under-served and/or
   decrease delivery rate to a group that is over-served. 

Specifically, when an ad has a chance to be shown to a user
  who is using one of Meta's platforms,
Meta's ad delivery system runs an ad auction between
  all ads targeting that user. 
In this auction,
  the ads compete based on their \emph{total value},
  which is calculated using
  the advertiser's bid,
  the \emph{estimated action rate}, which is how likely a user is to take the
  advertiser's desired action such as clicking on the ad,
  and \emph{ad quality score}, which estimates overall quality of the ad's
  content such as its image and text~\cite{vrscomplianceJune24}.
The total value is given by the following formula:
$
\textbf{Total Value} = \textbf{Advertiser Bid} \times \textbf{Estimated Action Rate}  + \textbf{Ad Quality Score}.
$
Importantly, all values in the formula, including advertiser bid, estimated action rate and ad quality score, are controlled by Meta's machine learning, and are not available to the advertiser.

VRS introduces a new parameter called \emph{VRS multiplier} that modifies the
  advertiser bid component to change the likelihood of an ad winning the
  auction~\cite{FacebookvsHUD2023TR}.
The direction of adjustment is chosen by a machine learning module trained on past data
 that takes as an input the latest privacy-protected measurement of variance,
  and produces either an \emph{adjust up} or \emph{adjust down} action that
  aims to shift the delivery ratios towards the eligible ratios.
The module does not receive individual-level demographic information such
   as the gender or race.
 Instead, it is trained to take as an input
  an embedding that summarizes the potential ad viewer
  along with the latest variance measurement
  to predict the most likely action
  that reduces variance for all demographic groups~\cite{Timmaraju_2023}. 
Few further specifics are known about the module, and importantly, the advertiser does not receive any information about the multiplier's or other influence of VRS on  their campaigns.
Meta's ad platform generally uses auto-bidding where advertisers
  specify a budget and the platform bids for each user on the advertiser's behalf; the bids and individual user costs are not reported to the advertiser.
The adjustments VRS makes and the resulting changes in costs are also not reported to the advertiser; and are thus completely opaque. 
In \autoref{sec:method1_exp_utiliery_user_adver},
  we show this implementation approach increases costs for advertisers,
  resulting in lower exposure to economic opportunity ads for users.

\subsection{External Verification of Compliance}
\label{sec:verification}

The settlement requires that a third-party entity serve as an external reviewer to confirm
  that Meta meets the compliance metrics~\cite{vrscomplianceJune24}.
The reviewer (currently, Guidehouse) is proposed and paid by Meta, but is subject to consent by the DoJ.

\begin{figure*}
  \centering
\includegraphics[width=0.8\linewidth]{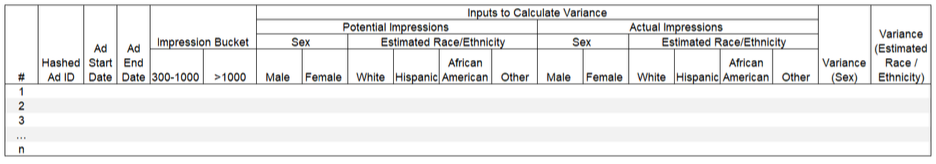}
    \caption{Meta VRS Compliance Metrics Reporting Schema~\cite{vrscomplianceJune24}}
     \label{fig:complianceschema} 
\end{figure*}

The external reviewer performs its analysis based on aggregate housing ad data that Meta reports to it every four months, using the data schema shown in \autoref{fig:complianceschema} (as documented by the reviewer~\cite{vrscomplianceJune24}).
Specifically, for each housing ad with 300+ impressions, identified using a hashed ad id, Meta includes:
  \emph{potential impressions}, i.e. 
  the number of impressions each demographic group
  in the eligible audience\footnote{Meta uses only a sample of the eligible audience for estimating variance and coverage 
  The external reviewer reports Meta's system has a target sample size of 6,000 users~\cite{vrscomplianceOct24}.
  }
  received in the last 30 days;
  \emph{actual impressions} broken down by demographic group;
  and Meta's calculation of \emph{variance} by gender and estimated race\footnote{In our description of VRS we omit details that relate to measures taken by Meta for privacy reasons, that include adding noise to the mechanism used to measure actual impressions and variance. We expand on these details in \autoref{sec:DP_vrs_overview}.}.

To verify compliance, the external reviewer simply computes the variance
  and coverage based on the aggregated data provided by Meta and using the variance formula given in \autoref{sec:settlement},
  and compares it with the variance and coverage metrics reported
  by Meta.
The reviewer also compares the achieved coverage against the targets
  agreed upon in the settlement (\autoref{tab:coverage}).

Notably, the reviewer has no means to get privileged access to Meta's internal data,
  and since the reviewer does not run test ads,
  the reviewer's ability to 
  independently verify accuracy of potential or actual impressions reported by Meta
  and the accuracy of their breakdown by demographics is limited.
Furthermore, the reviewer does not receive any information on costs. 
Finally, as we discuss in \autoref{sec:DP_vrs_overview}, the information reviewer receives is privacy-protected, which further interferes with its accuracy and verifiability.

While our work also operates without access to internal data,
  we suggest that our methods using test ads can strengthen
  an external audit.
Such tests are critical for providing a more independent and
  broader-scope verification of compliance that goes beyond
  simply verifying coverage computation formulas.

\subsection{Related Work}
  \label{sec:related_work}  
Prior to our work, the only type of external
  audit VRS's implementation has undergone, to our knowledge, is by the reviewer receiving 
  the periodic compliance reports mandated by the
  settlement~\cite{vrscomplianceJune24} (see \autoref{sec:verification}).
In addition, a 2023 article by a European non-profit, 
AlgorithmWatch,
  identified important information gaps in the compliance reports,
  called into question the scalability of VRS to other domains with risks of bias,
  and underscored the need for ``adversarial audits'' that are fully
  independent of Meta~\cite{vrsalgorithmwatch}.
In contrast, our work provides the first such fully independent and systematic critique
  of the settlement terms
  and VRS's implementation from the perspective of ability to mitigate discrimination
  and identifies further gaps in information needed for assessing the system. %
We also conduct the first black-box audit of VRS's impact on the
  delivery of real-world opportunity ads and their cost,
  filling some of the gaps and identify areas for improvement.

We next discuss related work on externally
  auditing discrimination in ad delivery,
  strategies for mitigating discrimination and
  their trade-offs with utility of ads for users and advertisers.

\paragraph{External Auditing of Discrimination in Ad Delivery}  
The methodology we develop for our experiments is inspired by prior
  audits that rely on paired ads to evaluate discrimination in ad delivery algorithms~\cite{Ali2019a,Imana2021,Imana2024}.
Discrimination in ad delivery occurs when a platform's algorithms
  show an ad disproportionately more (or less) often to members of certain
  demographic groups as a result of the platform-driven machine learning choices
  which determine which ads are displayed to users and
  how much to charge them~\cite{Sweeney2013, Lambrecht2016, Datta2015, Ali2019a}.
Such bias at the \emph{ad delivery} stage can emerge even when advertisers' \emph{ad targeting}
  is neutral—that is, when the advertiser chooses neutral targeting parameters
  that do not intentionally favor or exclude any demographic group.
Ali and Sapiezynski et al.~\cite{Ali2019a, Ali2019b} were the first to develop a
  paired ads methodology that isolates the role of platforms' algorithms from
  advertiser's targeting choices and other
  confounding factors to show ad delivery outcomes that are skewed
  by gender and race.
Subsequent studies~\cite{Imana2021, Imana2024} showed such skewed outcomes
  may be in violation of anti-discrimination laws for employment
  and education domains.
The methodology we develop in this work builds on the paired
  ads methodology, but focuses on  isolating the effect of VRS. 
In related work, Sapiezynski et al.~\cite{Sapiezynski2022} conducted
  an external audit of Meta’s Lookalike and Special Ad Audiences tools,
  uncovering persistent demographic bias despite the measures
  mandated by a 2019 settlement between Meta and the National Fair
  Housing Alliance~\cite{FacebookvsNFHAA}.
Their findings demonstrate that settlement commitments alone do not
  guarantee equitable delivery and underscore the need for
  independent, outcome-focused audits—a need that also motivates us
  to test whether VRS’s agreed upon compliance metrics
  lead to increased access to opportunities for users.

  \paragraph{Tradeoffs Between Fairness and Utility}  
  
Our evaluation of the tradeoff between VRS's fairness guarantees and utility
   for users and advertisers builds on prior body of work that study similar
   tradeoffs in both ad delivery and other algorithmic decision making
   systems~\cite{Friedler2019, Dwork2019, Hu2020, Kim2020, Rodolfa2021, Peysakhovich2023, Baumann2024, vladimirova2024fairjob}.
Closest to our study is the work by \citet{Baumann2024}
  that showed through simulations that
  fairness interventions in ad delivery can lead to a leveling down
  effect unless platforms
  explicitly share the cost of ensuring fairness.
Similarly, Hu and Chen~\cite{Hu2020} show enforcing
  group fairness metrics may not translate to improved outcomes
  to previously disadvantaged groups. 
Pesysakhovich et al.~\cite{Peysakhovich2023} show that fairness constraints in two-sided
  markets such as ad delivery can discourage advertiser participation,
  highlighting the trade-offs involved in designing effective fairness interventions.
Our research questions are motivated by these known tradeoffs
  between fairness and utility in machine learning.
Our work is the first to study these tradeoffs in the Meta-DoJ settlement  context
  and test how the VRS's current implementation affects utility for advertisers and users.

Balancing the interests of different stakeholders while achieving
  fairness is also an open research area in the broader field of recommender
  systems~\cite{Deldjoo2023Fairness, Stray2024}.
Our findings show that these trade-offs can cause certain VRS implementations, including the current one, 
  to inadvertently lower access to opportunities.
This result shows the need for platforms such as Meta to
  be transparent about how they manage these trade-offs.
   
\paragraph{Strategies for Mitigating Discrimination in Ad Delivery}

A number of approaches have been proposed for mitigation of
  discrimination in ad delivery~\cite{Celis2019, Dwork2019, Nasr2020, ilvento2020multi, chawla2022individual}.
Dwork and Ilvento~\cite{Dwork2019} demonstrate that fairness guarantees for
  individual components of a complex system, such as in ad delivery,
  do not necessarily translate into fairness for the entire system,
  and propose methods for combining seemingly unfair components
  to achieve fair outcomes.
Celis et al.~\cite{Celis2019} propose imposing fairness constraints on ad auctions
  to ensure balanced exposure across demographic groups.
Work by~\cite{Nasr2020} proposes an alternative approach that modifies the bids
  set by advertisers to mitigate discrimination without changing
  the underlying auction mechanism, and \cite{shen2021robust} looks for ad allocation mechanisms that do not hurt advertisers when some of them express diversity constraints.
While we do not propose a concrete and final solution for mitigating
  discrimination in ad delivery,
  we explore a budget-splitting approach that can lead
  towards an alternative to the VRS implementation chosen by Meta
  that is more effective
  at equitable delivery, transparent and explainable.

\section{Analysis of Meta/DoJ Settlement Terms}
\label{sec:design_critique}

We identify gaps in the settlement for mitigating discrimination in ad delivery:
  it focuses on impressions, not individuals, so it may not increase the number of distinct individuals reached by an ad;
   its coverage requirement treats all ads above a delivery impression threshold equally,
   so it allows for selective application of VRS to small ads impacting few users
   while continuing to use the standard, discriminatory, delivery algorithm for large ads;
   and its requirements can be satisfied by leveling down access
   to opportunities, so no demographic group may benefit from the fairness intervention.

\subsection{Focused on Impressions, not Individuals}
  \label{sec:imp_not_indi}

The first gap is that the settlement's compliance metrics
  are all defined in terms of impressions rather than individuals reached (\autoref{sec:settlement}).
Ad platforms typically use two types of metrics to evaluate the performance of an ad:
  \emph{impressions} and \emph{reach}.
Impressions represent the number of times a given ad was shown overall
  whereas reach represents to how many unique user accounts
  the ad was shown~\cite{ReachMetric}.
Thus, an ad delivery algorithm can meet the variance threshold by showing
  it repeatedly to the same individuals, 
  providing no increase in how many people see the opportunity.
As an example, according to the settlement metrics, an algorithm that shows an ad once to 100 unique men and shows it 100 times to one woman
  achieves zero variance for gender, and thus perfect equity, even though many more men are exposed to the opportunity ad.

\textbf{Recommendation:} Variance should be measured with respect to reach, i.e. the number of individuals from each demographic group to whom an ad is shown, rather than in terms of impressions received by members of each demographic group.

\subsection{Coverage and Selective Application}
  \label{sec:coverage_limitation}

The second gap is that the coverage requirement allows selective application of VRS
  to exclude large campaigns.
Coverage is defined as the fraction of housing ads for which VRS reduces
  variance below a certain threshold.
Meta has full leeway to choose the subset of ads for whom to meet the variance threshold.
In a hypothetical scenario,
  Meta might choose to meet the variance threshold for ads with
  small budgets or small audiences while allowing ads with large
  budgets or large audiences to fall into the (permitted) fraction
  that does not need to bound their variance.
This choice could result in ensuring fairness in delivery for a much smaller number of
  individuals and impressions than what the coverage metric alone might indicate. 
For example, ads with 1k and 1M impressions count equally to the coverage target,
  but the latter affects $1000\times$ more people.

We illustrate the magnitude of spend and impressions that could be excluded from the fairness intervention given this slack in the settlement using public data Meta provides on political advertising budgets and reach\footnote{Our illustration assumes distributions of impressions and spending for ads for economic opportunities follows similar patterns to those of political ads. This assumption is  consistent with data published by Meta~\cite{wernerfelt2022estimating}.}.
We obtain a representative sample of 32,867 political ads ran in the US in 2024 using the Meta Ad Library and a methodology developed in prior work~\cite{nagaraj2023discrimination}.
We exclude ads that received fewer than 300 impressions as settlement terms do not apply to such ads.
We find the distributions of spend and impressions for the sample of ads
  follows a power-law relationship~\cite{Manning2008IR, powerlaw},
  where most of the ads
  spend a small amount and receive relatively few impressions, 
  but a few ads spend a large amount
  and receive a very large number of impressions
  (see~\autoref{sec:meta_ad_dist} for figures).

To quantify the implications of this power-law relationship under selective application of VRS,
  we examine the actual guarantees the compliance metrics provide
  depending on which ad campaigns are selected and omitted from variance reduction.
Specifically, among the 32,967 ads that together received 1.3 billion impressions,
  we compare the effect of selectively excluding the largest campaigns from variance reduction
  with the effect of randomly excluding campaigns.

Consider the 81\% coverage requirement for race at 10\% variance threshold (from \autoref{tab:coverage}),
  i.e. 19\% of ad campaigns are permitted to be excluded from
  variance reduction.
If the platform chooses the largest 19\% of the ads,
  then approximately 1.03 billion impressions (78.9\%) are not subject to fairness intervention.
In contrast, if the platform chooses a random 19\% of the ads, then
  (empirically, taking the mean of the number of impressions in 100 such experiments),
   only 250 million impressions (19\%) would be excluded from the fairness intervention.
Similarly, exclusion of the largest 39\%
  of ads (based on 61\% coverage requirement at 5\% threshold for variance by race)
  would exclude approximately 1.19 billion impressions (90.3\%) from the fairness intervention,
  again far more than the 517 million impressions (39\%) that would be excluded
  using random selection.
We observe similar trends -- that selective application allows exclusion of a far greater percentage of impressions than the coverage metrics implicitly imply, for all the coverage requirements.

\textbf{Recommendation 1:}
We recommend enforcing the
  coverage requirement within stratified tiers based on
  audience size and ad spend levels to reduce the uncertainty in how broadly VRS applies.
One starting point could be the categories of ``small''
  and ``large'' advertisers that Meta already uses internally~\cite{wernerfelt2022estimating}.

\textbf{Recommendation 2:}
We recommend for the external reviewer to check for selective application of VRS
  by conducting an analysis of the distribution of spend and impressions
  for ads that meet and do not meet the variance reduction thresholds.
  
\subsection{Risk of Leveling Down}
  \label{sec:critique_level_down}

The third drawback of the settlement is that it is possible to satisfy the
  compliance metrics
  by leveling down access to opportunities.
Leveling down is defined as achieving fairness by
 bringing the performance for better performing groups down to the level of worse performing groups~\cite{Mittelstadt2023}.
Recent work~\cite{Baumann2024} has shown through simulations on simplified models that fairness interventions in ad delivery specifically run that risk, unless one explicitly constrains the space of solutions to those where the total number of ad impressions does not decrease.
VRS's compliance metrics and the implementation
  have no such constraint.
In \autoref{sec:method1_exp_utiliery_user_adver},
  we run experiments with real ads that show
  how this limitation can lead to leveling down in practice.

\textbf{Recommendation:} 
We recommend future efforts to regulate ad delivery algorithms explicitly address the risk of leveling down.
\section{A More Complete Independent Evaluation of VRS's Performance}
\label{sec:method1}

We next present our external evaluation of the performance of VRS implementation by Meta in the delivery of real-world ads.
Our findings reveal that while VRS reduces variance as measured by its compliance metrics,
  it increases cost to advertisers,
  decreasing exposure of opportunity ads to recipients as a result.
When we subsequently refer to VRS, we mean the specific implementation Meta has chosen for the settlement compliance.

\subsection{Methodology}
  \label{sec:method1_isolating_vrs}

Our black-box methodology isolates the effect of VRS by running the
  same ad twice: once with and once without VRS. We then evaluate the relative
  performance of the ads by the demographic attributes of interest,
  such as race and gender.
We run both copies with the same targeting parameters,
 including the ad creative, the budget, and demographic makeup of audiences,
 so that the only difference between the two ad configurations
 is the presence or absence of VRS application to their delivery.
This approach builds on prior work that used paired ads for isolating the role of the ad delivery algorithm from other factors when auditing for discrimination~\cite{Ali2019a, Ali2019b, Imana2021, Imana2024},
  but is the first to use it to evaluate a system built to mitigate discrimination.
  
\subsubsection{Isolating the Effect of VRS}
  \label{sec:enabled_disable_vrs}
To measure the effect of VRS,
  we enable or not enable VRS for each of our ad campaigns
  by declaring them or not declaring them as belonging to the special ad category of housing,
  the only category for which Meta is legally required to reduce variance.
We infer from Meta's and DoJ's public statements that
  VRS is automatically enabled for any ad
  an advertiser self-declares as housing~\cite{FacebookvsHUD2, FacebookvsHUD2023}.
We use non-housing ads in our experiments instead of housing ads because housing ads are required to be declared as such, and not doing so risks being rejected from running the ad and future ads.
To our knowledge, Meta does not reject non-housing ads that are labeled as housing ads.
Our declaration of a non-housing ad as a housing ad does not risk user harm,
  as it triggers stricter fairness constraints in delivery per VRS.

 We use ads whose delivery may be skewed towards
  a particular demographic group when VRS is not active
  to clearly see the effect when VRS is enabled.
The six types of ad creatives we use and their rationale are summarized in \autoref{tab:ads_summary}.
First,
  we use two non-opportunity ads that are stereotypically
   associated with a particular demographic group:    an ad for a hair product that is stereotypically associated with Black women
   and an ad for golfing that is stereotypically associated with White men.
We expect their delivery to be skewed towards users from those groups
    when VRS is not enabled.
Second, we examine two ads for education opportunities,
  an economic opportunity that prior research has shown is
  susceptible to discriminatory ad delivery~\cite{Imana2024}.
Third, we analyze two ads for insurance and financial products,
  which are domains not studied in previous research,
  but also have risks of discrimination and are
  planned to be added to categories of ads for which Meta voluntarily applies VRS 
  beginning in January 2025~\cite{FacebookCreditExpansion}.

We measure variance by race and gender for both the VRS and no-VRS ads,
  and test whether enabling VRS reduces variance.
From the racial and gender groups in VRS's scope (summarized in \autoref{sec:settlement}),
  we measure variance using both gender groups and the two largest racial groups
  most well-represented in our source dataset for building ad audiences: Black and White.

\newcolumntype{L}[1]{>{\raggedright\arraybackslash}p{#1}}
\begin{table}
\centering
\begin{tabular}{|L{0.6cm}|L{2.1cm}|L{4.6cm}|}
\hline
\textbf{ID} & \textbf{Ad creative} & \textbf{Description}\\ \hline
HA & Hair product  & Stereotypically skew: Black, Women \\ \hline
GA & Golfing & Stereotypically skew: White, Men \\ \hline
EA & Education (ASU) & Evidence of bias from~\cite{Imana2024} \\ \hline
EB & Education (CSU) & Evidence of bias from~\cite{Imana2024} \\ \hline
IA & Insurance & VRS may apply starting 2025 \\ \hline
FA & Financial & VRS may apply starting 2025 \\ \hline
\end{tabular}
\caption{List of categories of ad creatives we use in our experiments. Example screenshots are given in~\autoref{fig:ad_image_all}.}
\label{tab:ads_summary}
\end{table}

\begin{figure}
\centering
\begin{subfigure}[T]{0.43\linewidth}
    \centering
    \includegraphics[width=\linewidth]{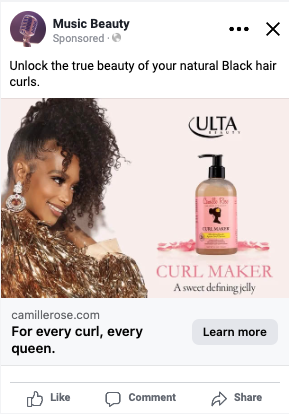}
    \caption{Hair product ad}
     \label{fig:ad_image_hair} 
\end{subfigure}
 \begin{subfigure}[T]{0.415\linewidth}
    \centering
    \includegraphics[width=\linewidth]{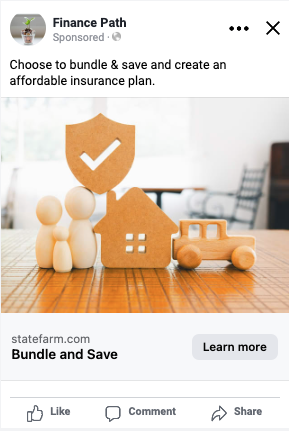}
    \caption[b]{Insurance ad}
    \label{fig:ad_image_insurance}
 \end{subfigure}
 \caption{Example screenshots of our ad creatives.}
 \label{fig:ad_image_all}
\end{figure}

\subsubsection{Building Ad Audience}
  \label{sec:ad_audience}
 
For our experiments,
  we use audiences that are demographically balanced by
  both race and gender.
We specify the audiences using Meta's Custom Audience feature
  that allows us to upload a list of individuals' personal information
  such as their names and location, which then Meta matches
  to real user accounts.
We select individuals using North Carolina's (NC) voter
  dataset, a public source of location, gender and race of
  individuals~\cite{VoterDataNC}.
  
Since Meta does not report race of ad recipients, for experiments where we study variance by race, 
   we build our ad audiences in a way that allows
  us to infer race from the location of ad recipients,
  following the approach
  in prior external audits of ad delivery~\cite{Ali2019a, Imana2021, Imana2024}
  (see \autoref{sec:loc_proxy} for details).
In each audience partition,
  we include a list of 30k individuals balanced by both
  race and gender:
  7.5k White-male, 7.5k Black-male,
  7.5k White-female, and 7.5k Black-female individuals.
 
\subsubsection{Other Campaign Parameters}

We run both the VRS-enabled and no-VRS copies of an ad for a 24-hour period
  with a total budget of \$20 per ad.
\footnote{As in prior work focused on studying ad delivery,
 we use ``Traffic'' objective for all ads,
  which optimizes delivery for increasing traffic, i.e. clicks, for websites
  our ads link to. In order for ads not to waste recipients' time, each ad links to real websites where users
  can learn more about the product or opportunity we advertise.}
We find our chosen audience sizes, campaign durations and budgets are
  large enough to generate sufficient ad impressions
  for our analyses.
In our experiments in \autoref{sec:experiments_method1}, we observe that VRS begins to take effect after a few hours
  of delivery.
Meta’s settlement requires VRS to reduce variance for all ads that receive at
  least 300 impressions (see \autoref{sec:settlement}),
  and our ads generate approximately 1,500 impressions
  on average (all get at least 1,100 impression)—well above this threshold.
    
To avoid the two ads with the same visual elements competing
  for the same set of users,
  we run the VRS-enabled and no-VRS ads on separate audience partitions.
This step differs from our prior work where we ran ads for different 
  opportunities concurrently targeting the exact same audience~\cite{Imana2021, Imana2024},
  so we could explicitly see which ad the ad delivery algorithm favors.
Here we run identical ads, varying only the status of VRS,
  so we run on different audiences to avoid self-competition.
Using different audiences introduces a potentially confounding
  factor—different competition from other advertisers—a control present
  in prior work but incompatible with our methodology because we
  avoid self-competition.
We repeat each experiment on three randomly sampled audiences
  to ensure reproducibility.

\subsubsection{Experiment Setup}
  \label{sec:exp_setup}

Each experiment we run consists of a pair of a VRS-enabled and a no-VRS ad
  that are otherwise identical.
We evaluate variance by both race and gender.
To validate our results with multiple replications, 
  we reproduce each experiment on three disjoint audience partitions
  randomly sampled from the NC voter dataset.
Using the six ad creatives summarized in \autoref{tab:ads_summary},
  two demographic attributes,
  and three replications,
  we run a total of $N=36$ experiments of paired no-VRS ads
  and VRS-enabled ads.
  
\subsubsection{Estimating Eligible Ratio}
  \label{sec:eligible_ratio_method}
We estimate the eligible ratio used by Meta for our ads by
  taking the mean of all delivery ratios we
  observe for each attribute across all
  VRS-enabled ads we run.
As defined in \autoref{sec:settlement},
  VRS uses the eligible ratio as a baseline for reducing
  variance.
As external auditors,
  we do not have access to data about past ad impression
  breakdowns by demographic group,
  which is the data used by Meta to determine the eligible ratio.
Assuming that VRS works as intended and by the law of large numbers,
  the mean of the delivery ratios is a reasonable estimate.
We note, however, that our estimate of eligible ratio does not account for
  the up to 10\% variation that the compliance metric allows between
  eligible ratio and delivery ratio, and the noise from VRS's reliance on
  BISG and privacy-protecting measures (\autoref{sec:differential_privacy_in_VRS}).
We do not see a path to a more precise estimate with only external information. 

\subsection{Experiments}
  \label{sec:experiments_method1}
We next apply our methodology to real-world ads
  to support our two key claims:
  VRS does reduce variance with respect to the metrics agreed upon in the settlement (\autoref{sec:method1_exp_eligible});
  and the impact of enabling VRS is higher per-ad costs for advertisers and therefore
    fewer opportunity ads shown to recipients for a given budget (\autoref{sec:method1_exp_utiliery_user_adver}).

\subsubsection{Does VRS Reduce Variance with Respect to Eligible Ratio?}
  \label{sec:method1_exp_eligible}

We confirm that VRS does reduce variance with
  respect to our estimate of the eligible ratios, i.e. with respect to the compliance metrics agreed upon in the settlement
  for delivery of housing ads.
Our estimate of eligible ratios using heuristic from  \autoref{sec:eligible_ratio_method}
  is 0.42, 0.58, 0.45, and 0.55 for
  Black, White, male and female groups, respectively.\footnote{In \autoref{sec:skew_in_delivery_ratio}, we show that the non-balanced delivery ratio
  even though the Custom Audiences we target are balanced 50:50
  can be explained by differences in rates at which our audience matches with real accounts.}

\begin{figure}
  \centering
     \includegraphics[width=0.38\textwidth]{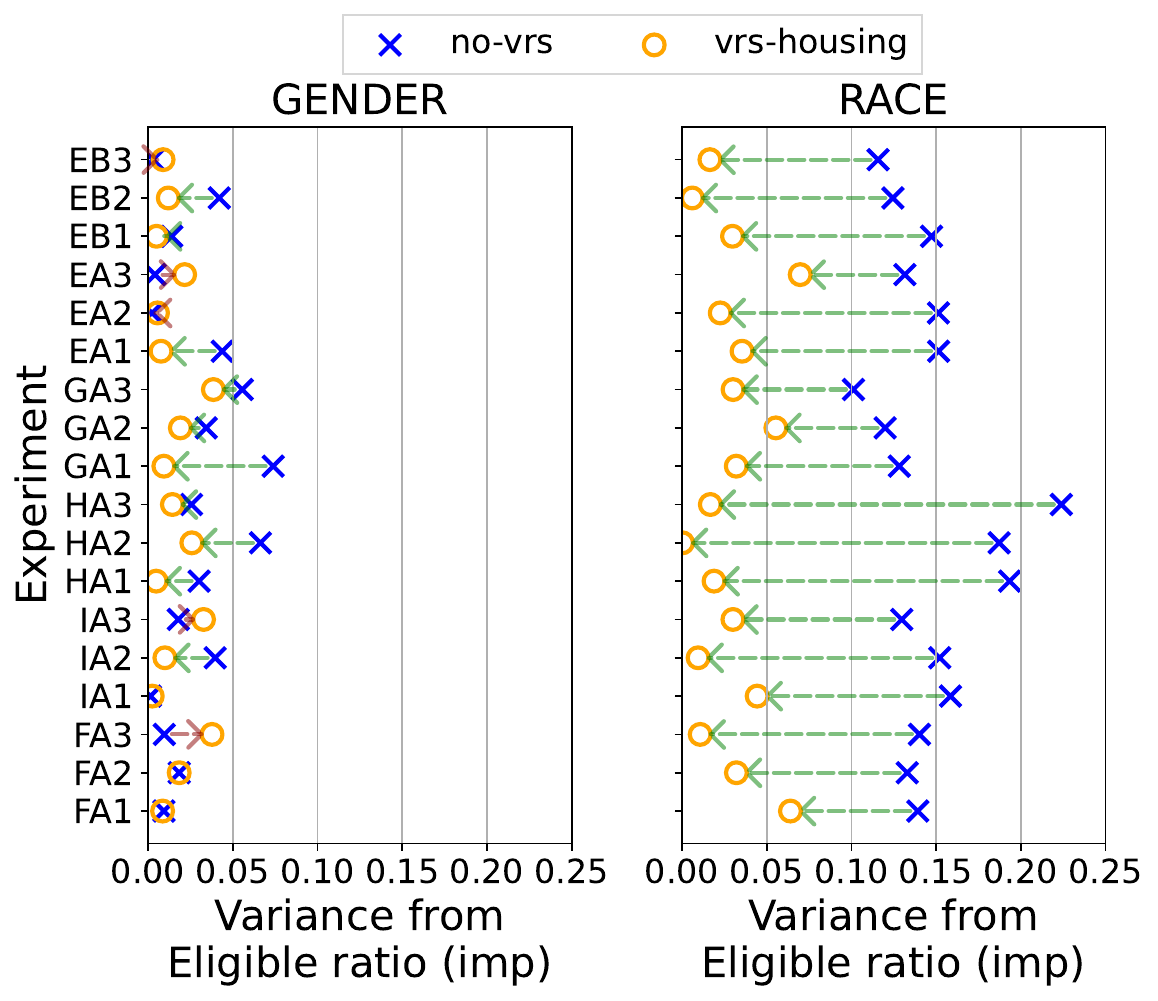}
        \caption{Comparison of variance with and without VRS. VRS generally reduces variance compared to the no-VRS case.}
        \label{fig:vrs_vs_novrs_variance-eligible}
\end{figure}

Using our estimates of the eligible ratios,
\autoref{fig:vrs_vs_novrs_variance-eligible} compares the variance
  of all $N=36$ pairs of VRS and no-VRS ads.
The orange dots (vrs-housing) and the blue cross-marks (no-vrs)
  show the level of variance when VRS is enabled and disabled, respectively.
The green arrows indicate that, compared to no-VRS,
  enabling VRS reduces variance.
A red arrow indicates VRS increases the variance instead.

In the left figure for gender,
  variance is less than 5\% even without VRS in 15 out of 18 cases,
  but enabling VRS further reduces it in some cases. 
In the right figure for race,
  we see variance is more than 10\% without VRS
  in all 18 cases.
VRS reduces variance to less than 10\% in all cases,
   bringing it down to less than 5\% in 15 of the 18 cases.
This result is the first in-the-wild verification
  that VRS reduces variance with respect to the compliance metrics
  specified in the settlement.

\subsubsection{Evaluating VRS's Performance on Employment and Credit Ads}
  \label{sec:credit_and_job}

\begin{figure}
\centering
\begin{subfigure}[T]{0.68\linewidth}
    \centering
    \includegraphics[width=\linewidth]{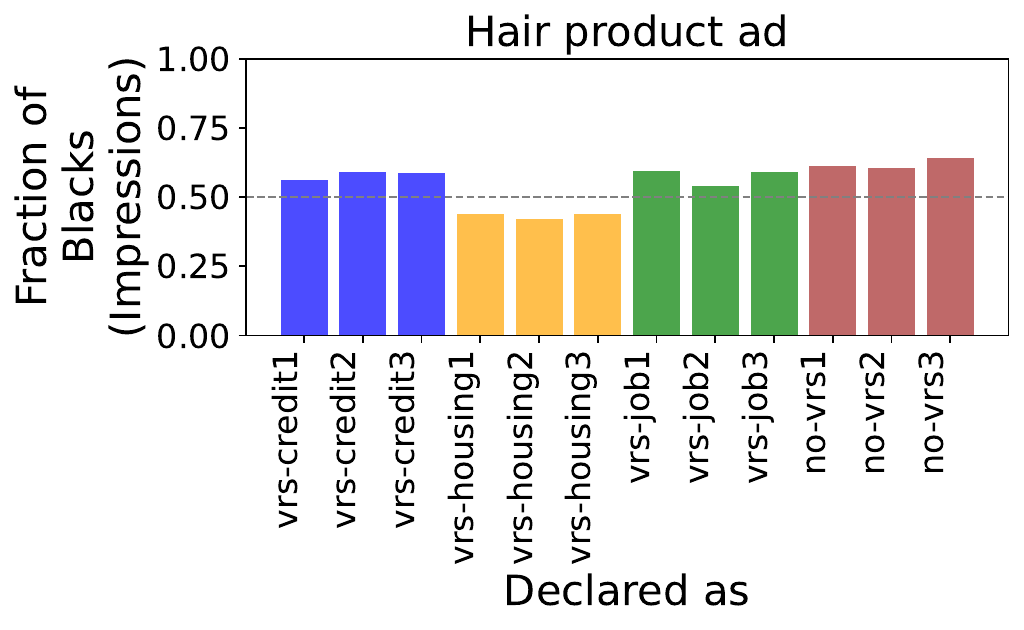}
    \caption{Demographic breakdown of ad impressions}
     \label{fig:heccomp_b_frac} 
\end{subfigure}\hfil
\begin{subfigure}[T]{0.68\linewidth}
    \centering
    \includegraphics[width=\linewidth]{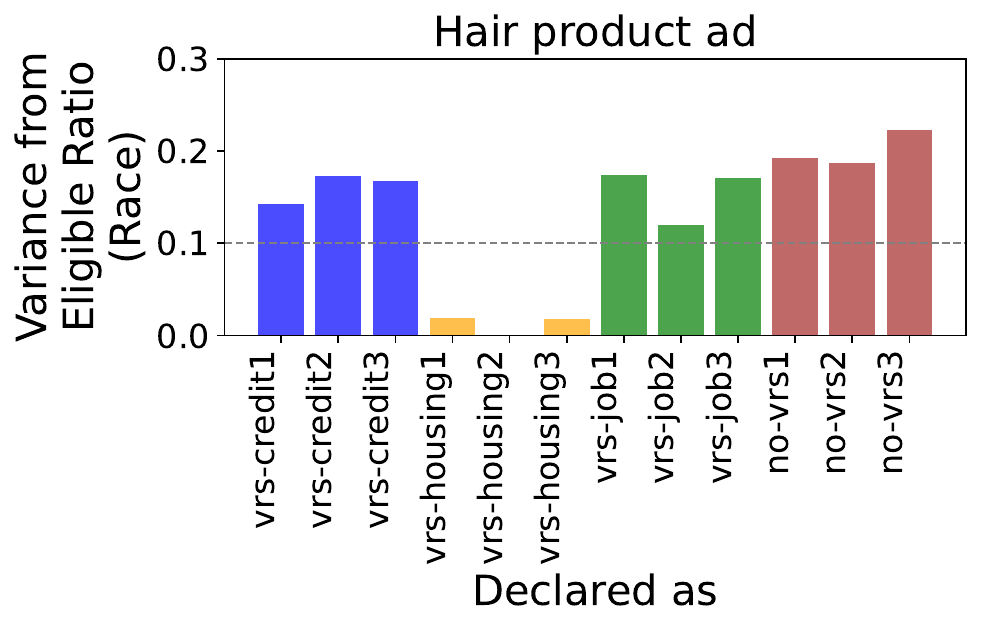}
    \caption[b]{Variance from estimated eligible ratio}
    \label{fig:heccomp_variance}
 \end{subfigure}\hfil
 \caption{Comparison of VRS's performance on housing, the only domain mandated by settlement, with employment and credit, domains for which Meta voluntarily deployed VRS.}
 \label{fig:heccomp}
\end{figure}

\basi{}
We next evaluate whether the reduced variance we observed for housing ads
  in the previous section also occurs for employment and credit ads — two special ad
  categories to which Meta has voluntarily expanded its application of VRS.
Because Meta states employment and credit domains use the same VRS framework,
  we expect VRS to operate similarly and reduce variance in all three domains. 
However, unlike housing, Meta is not legally required to meet the same variance
  thresholds or coverage levels in employment and credit,
  potentially allowing for a higher variance tolerance or
  lower coverage in those domains.

For this evaluation, we run additional experiments
  where we enable VRS by declaring an ad as a credit or an employment ad,
  instead of housing.
We then compare the outcome with the no-VRS and VRS-housing ads
  we ran in \autoref{sec:method1_exp_eligible} where we enabled VRS
  by declaring an ad as a housing ad.
We conduct this evaluation using
  the hair product ad which we expect to skew towards
  Black women when VRS is not enabled.
We reproduce all ad campaigns on three different audiences.
  
\autoref{fig:heccomp_b_frac} shows the outcome of VRS for all three domains: housing, credit and employment.
We compare the fraction of Black users each type of VRS-enabled ad was shown to.
The horizontal dotted line indicates 50\% delivery to Black users.
We see that an ad declared as housing is delivered to approximately 43-44\% Black
  users in all three repetitions,
  consistent with the overall trend of delivery ratio we observed VRS achieves for race
  in \autoref{sec:method1_exp_eligible}.
However, the ads declared as credit and employment 
  are delivered to 55-60\% Black users,
  an outcome that is closer to the 60-65\% Black outcome we see
  for the no-VRS case.

We further illustrate that the coverage and variance metrics Meta applies to its voluntary deployment of VRS to its credit and employment ads are more lenient than those applied to housing ads in \autoref{fig:heccomp_variance},
  where we compare the variance for all the ads using as a baseline
  the eligible ratio we estimated in \autoref{sec:method1_exp_eligible}.
The horizontal dotted line represents the maximum
  10\% variance threshold that is allowed for housing ads
  by the settlement.
This figure evidently shows VRS reduces variance to below
  10\% for the ads declared as housing,
  but not for the ads declared as credit or employment.
The variance for those two categories remains above 10\%
  and is comparable to the delivery outcome without VRS.

These findings indicate that the voluntary expansion of VRS
  to employment and credit domains, while a commendable initiative,
   does not match the implicit expectation
  that it performs in the same way as in the housing domain.
Our results add to evidence that self-regulation is rarely sufficient~\cite{Ali2019a, Imana2021, Imana2024},
  and support our recommendation that the target coverage and variance metrics
  should be released by the platform in order to meaningfully claim
  extension of compliance to other domains.

\subsubsection{VRS Reduces Utility for Users and Advertisers}
  \label{sec:method1_exp_utiliery_user_adver}

\begin{figure*}
\centering
\begin{subfigure}[T]{0.21\linewidth}
    \centering
    \includegraphics[width=\linewidth]{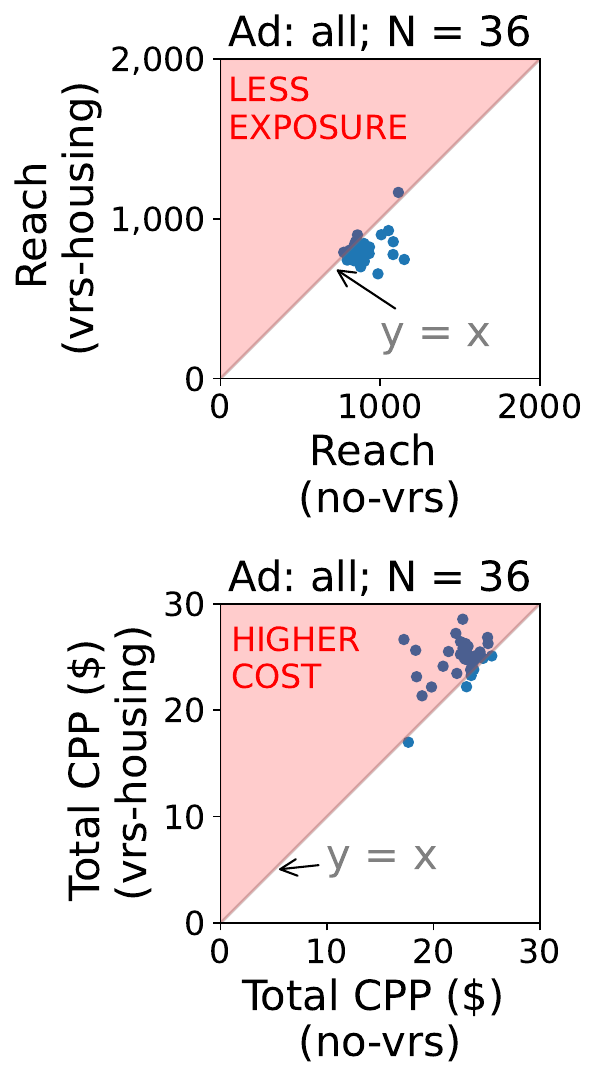}
    \caption{Total reach and CPP}
     \label{fig:scatter} 
\end{subfigure}\hfil
\begin{subfigure}[T]{0.28\linewidth}
    \centering
    \includegraphics[width=\linewidth]{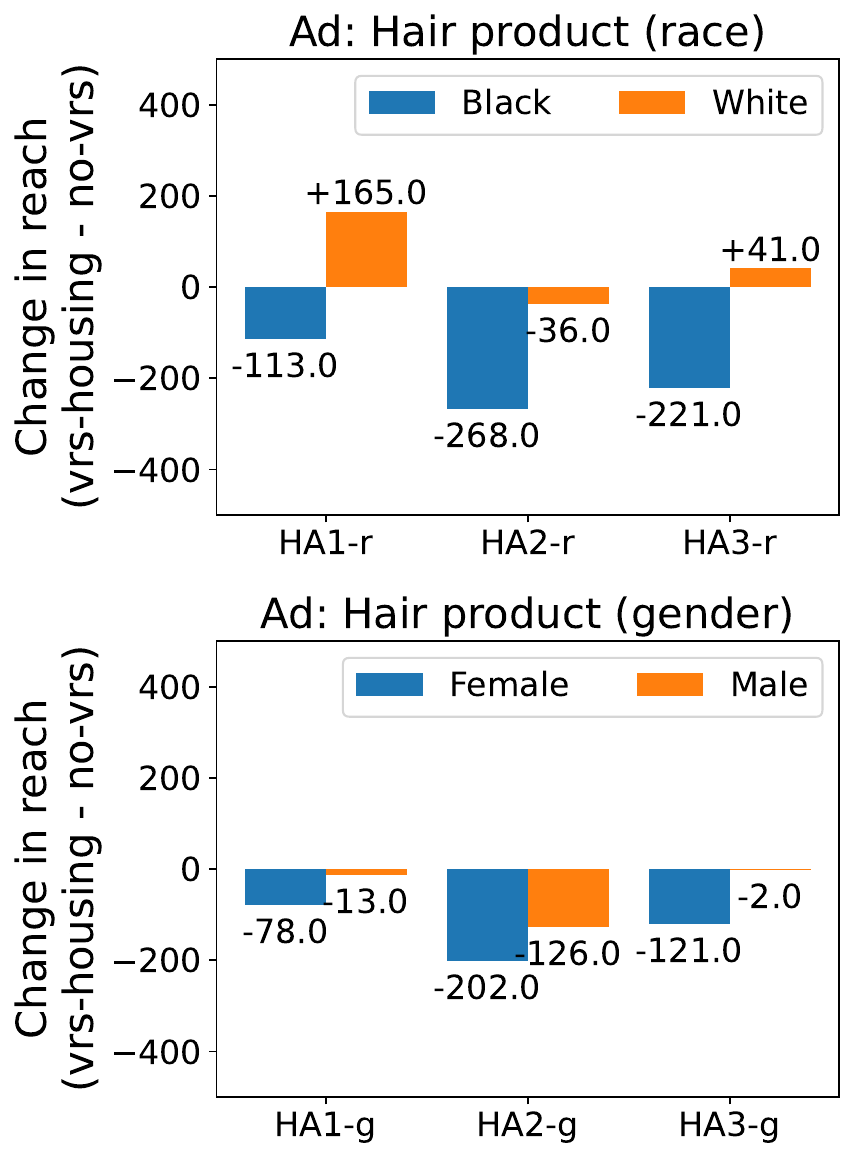}
    \caption[b]{Change in reach per group.}
    \label{fig:per_group_impressions}
 \end{subfigure}\hfil
 \begin{subfigure}[T]{0.28\linewidth}
    \centering
    \includegraphics[width=\linewidth]{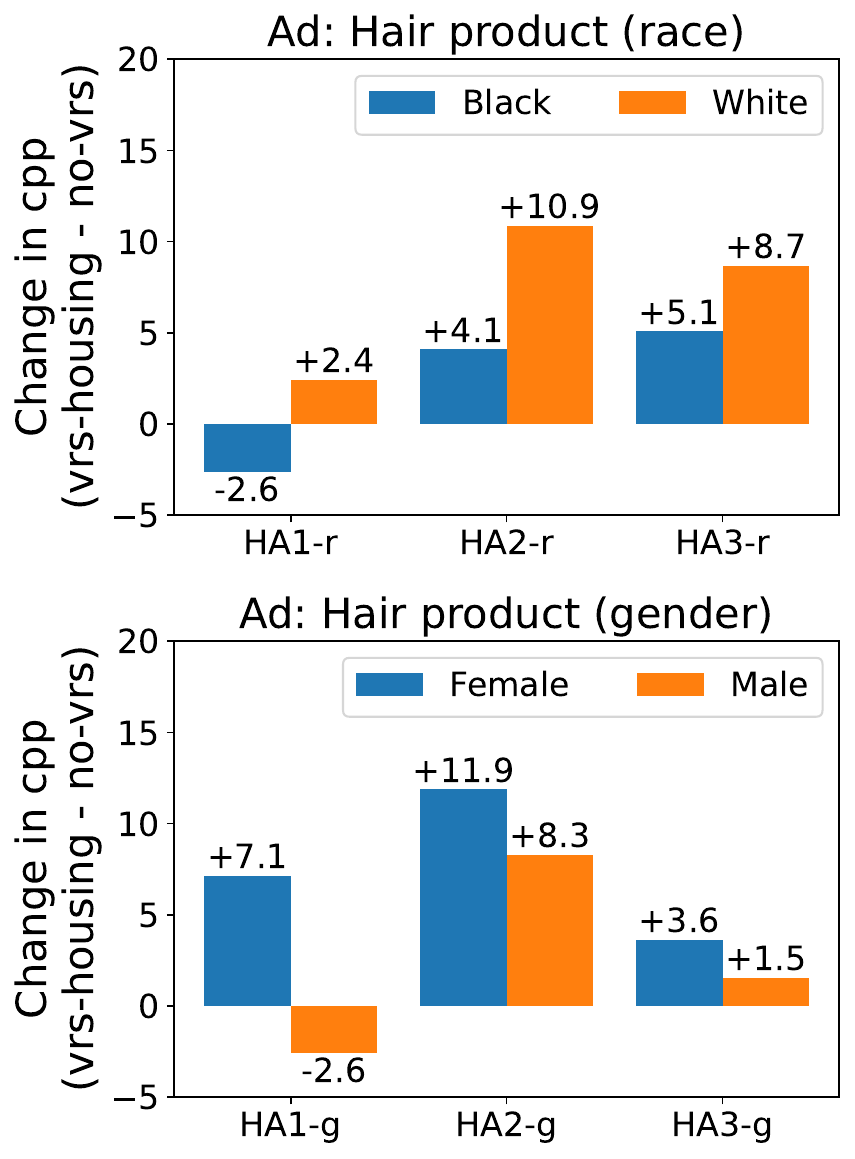}
    \caption[b]{Change in CPP per group.}
    \label{fig:vrs_change_in_cost}
 \end{subfigure}
 \label{fig:vrs_vs_novrs_cost}
 \caption{Comparison of reach metric and cost per 1,000 reach (CPP) with and without VRS.}
\end{figure*}

We next evaluate how VRS affects
  exposure to opportunity ads for users
  and costs for advertisers.
We measure 
the number of unique people
  an opportunity ad is shown to across different demographic groups (i.e. ad's reach).
We also test whether VRS results in the leveling-down effect
  (hypothesized in \autoref{sec:critique_level_down}),
  where lower variance is achieved by decreasing ad exposure for the advantaged
  group without benefiting the previously disadvantaged group.
We measure utility for advertisers in terms of cost per reaching
  1,000 unique ad recipients of a certain demographic (CPP or ``Cost per Purchase''), which captures how advertisers are affected by VRS.
We find that the cost of VRS is passed on to advertisers and the VRS implementation does not necessarily lead to a greater exposure of opportunities to users.

The scatterplots in \autoref{fig:scatter} comparing the reach and
  CPP for the same ad run with and without VRS enabled reflect the results of 36 experiments.
Each point corresponds to a paired ad experiment, with its reach (resp. CPP) for the no-VRS version on the $x$-axis, and reach (resp. CPP) for the VRS version on the $y$-axis.
In the top figure,
 the majority of the points in the scatterplot lie below the diagonal line
  ($y = x$), indicating that fewer people receive the ad when VRS is enabled.
Across all the paired ads, enabling VRS reduces mean reach by 9.82\%.
In the bottom figure, the majority of the points lie above the diagonal line, indicating that the cost increases when VRS is enabled.
Enabling VRS increases CPP by 12.02\% on average
  across all experiments.
Taken together, these results
  demonstrate that enabling VRS results in fewer exposures
  to opportunities for individuals, with each ad recipient costing advertisers more.

We next consider how the decrease in exposure and increase in cost
  is distributed among different demographic groups.  
For this evaluation,
  we look at the hair-product ad creative whose delivery we expect
  to be skewed towards Black women when VRS is not enabled across 6 paired experiments on disjoint audiences.
The bar chart in \autoref{fig:per_group_impressions}
  shows the effect of enabling VRS on the number of users
  reached by the ad from each demographic group, with the top figure giving the breakdown by race, and the bottom one by gender.
As expected based on the creative we use,
  VRS reduces the number of Black users and the number of women who see the ad,
  compared to its no-VRS option.
Furthermore, 
we do see the leveling-down effect in 4/6 cases (HA2-r,  HA1-g, HA2-g, and HA3-g), 
  where both groups see a decrease in exposure.
In another case (HA3-r),
  one group experiences a substantial reduction in
  ad exposure while the other does not receive a
  corresponding increase.

\autoref{fig:vrs_change_in_cost} presents the results of the same 6 paired experiments analyzed
  from the perspective of the cost to reach users from each group.
CPP consistently and significantly rises with enabling VRS (except in HA1-r and HA1-g),
  showing Meta pushes the cost of achieving fairness to advertisers.
Overall, these results demonstrate that VRS achieves lower variance
  by making impressions scarcer and, therefore, more expensive
  for advertisers.

\textbf{Recommendation: }
We suggest Meta explores alternative strategies
  that do not pass the cost of compliance fully to users and advertisers.
One way is to add a constraint on the number of impressions the VRS should achieve -- non-decreasing compared to the non-VRS version, as discussed in \autoref{sec:critique_level_down}.
Another strategy is to modify the Total Value computation or its use in price-setting in opportunity ads, so as to distribute the cost between the platform and advertisers.

\section{VRS is a Suboptimal Implementation of Settlement}
\label{sec:method2}

We next demonstrate reduced utility for users and advertisers shown in the previous section
  is an artifact of the specific implementation for VRS chosen by Meta,
  and not an inherent consequence of the settlement goals.
We show this by experimentally comparing the outcomes of VRS with
  a simple alternative approach aimed at reaching a demographically balanced audience:
  splitting an ad's total budget equally among all targeted demographic groups,
  and then running separate ad campaigns for each group\footnote{A more sophisticated version of this approach has been pioneered and empirically tested by \cite{gelauff2020advertising} prior to the Meta-DoJ settlement.}.
Based on the experiments in \autoref{sec:experiments_method2},
  we show this alternative approach outperforms VRS by
  increasing exposure to opportunities for all groups and
  reducing cost to advertisers, compared to the VRS-enabled run.
We caution that splitting the budget equally does not necessarily guarantee equal
  impressions; however, since even this simple approach outperforms VRS, we conclude that VRS is
  suboptimal.
  
\subsection{Methodology}
  \label{sec:budget_splitting}

We explore an approach where
  we manually split a campaign budget
  evenly among demographic groups instead of
  relying on VRS to automatically and implicitly determine spending on each group.
We run separate ad campaigns for each demographic group with
  an equal share of the total budget.
  We focus on both gender groups and the two largest racial groups
  most well-represented in our audience dataset: Black and White.
Consequently, we split the total budget among the following
  four subgroups: White males, White females, Black males, Black females.
We then run four separate ad campaigns targeting each subgroup with a quarter
  of the total budget.
We set the total budget to \$20, so each subgroup receives a budget of \$5.
We additionally run a single VRS-enabled campaign with the full \$20 budget,
targeting the combined audiences.
We aggregate the results across the four split campaigns and compare
  the outcome with that of the VRS-enabled campaign
  along the following metrics: variance, reach and cost for advertisers.
Other parts of the methodology, such as the ad creatives and audience sources, 
  follow our first methodology described in \autoref{sec:method1_isolating_vrs}.

\subsection{Experiments on VRS vs Budget-Splitting}
  \label{sec:experiments_method2}

Using the six different ad creatives in \autoref{tab:ads_summary},
   two demographic attributes (race and gender),
   and replication on three different audience partitions,
   we run a total of $N=36$ experiments comparing the outcome
   of ad delivery with VRS and with budget-splitting.

\begin{figure*}
\centering
\begin{subfigure}[T]{0.18\linewidth}
    \centering
    \includegraphics[width=\linewidth]{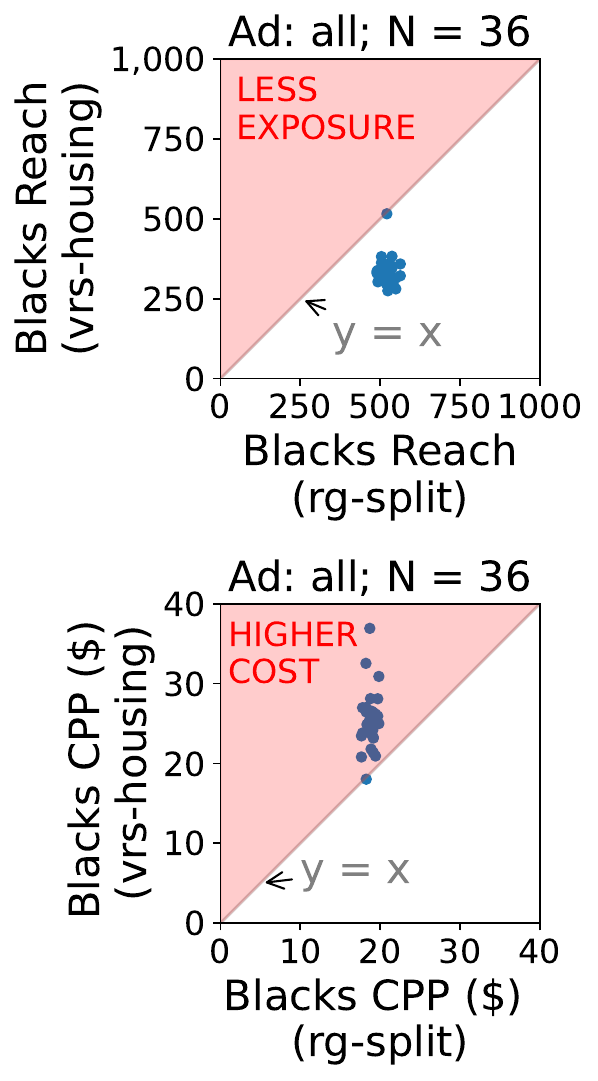}
    \caption{Black}
     \label{fig:vrs_vs_split_by_both_imp_cpm_b} 
\end{subfigure}\hfil
\begin{subfigure}[T]{0.18\linewidth}
    \centering
    \includegraphics[width=\linewidth]{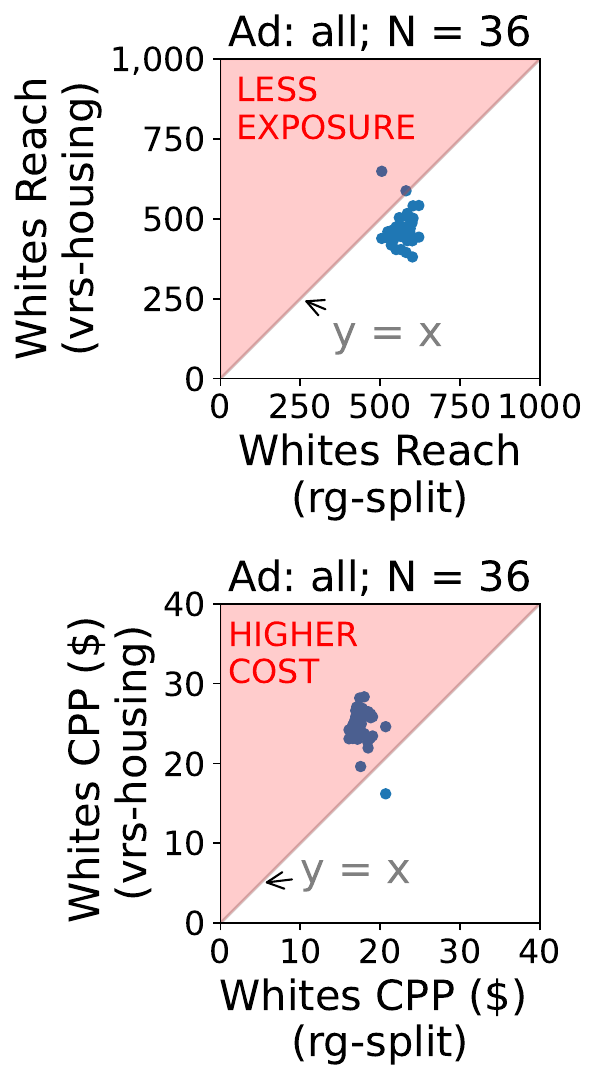}
    \caption[b]{White}
    \label{fig:vrs_vs_split_by_both_imp_cpm_w}
 \end{subfigure}\hfil
 \begin{subfigure}[T]{0.18\linewidth}
    \centering
    \includegraphics[width=\linewidth]{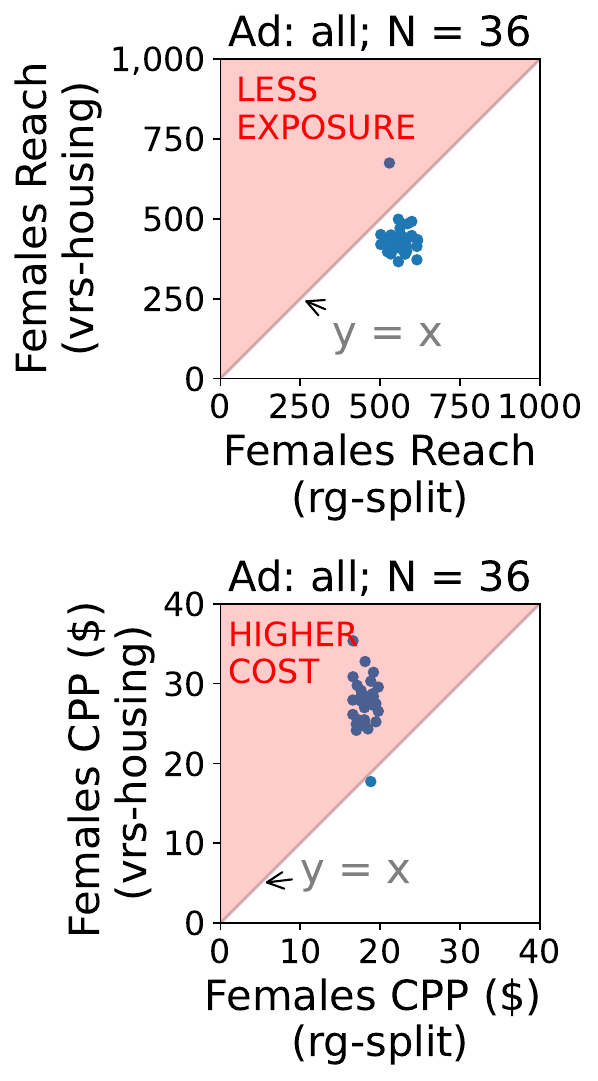}
    \caption[b]{Female}
    \label{fig:vrs_vs_split_by_both_imp_cpm_f}
 \end{subfigure}\hfil
 \begin{subfigure}[T]{0.18\linewidth}
    \centering
    \includegraphics[width=\linewidth]{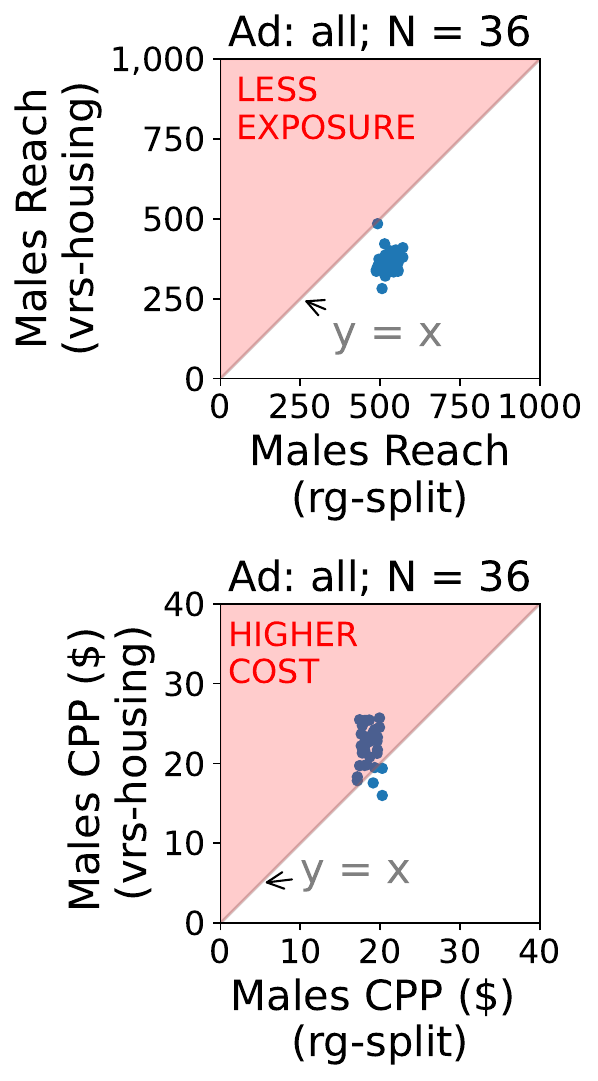}
    \caption[b]{Male}
    \label{fig:vrs_vs_split_by_both_imp_cpm_m}
 \end{subfigure}
 \caption{Reach and CPP by demographic group. The top figures show, compared to VRS,  the budget-splitting method increases impressions for all groups. The bottom figures mirror this result by showing VRS as a result has higher CPP.}
 \label{fig:vrs_vs_split_by_both_imp_cpm}
\end{figure*}

Across all demographic groups we
  consider, the budget-splitting approach
  outperforms VRS by increasing
  reach for all groups.
Top row of \autoref{fig:vrs_vs_split_by_both_imp_cpm} illustrates this result, by presenting ad reach achieved by the 36 pairs of ads when using the budget-splitting strategy ($x$-axis) vs. the VRS strategy ($y$-axis).
For Black, White, Female and Male users, the vast majority of the points are below the $y=x$ line, demonstrating that the budget-splitting method gives more exposure to that demographic group than the VRS method, while keeping the overall budget fixed. 
The increase in reach across the board shows
  budget-splitting achieves a categorically better outcome than VRS in terms
  of increasing access for the economic opportunities advertised.
Given we use the same total budget for all campaigns,
the reduced reach of VRS compared to budget-splitting 
  implies VRS also has a larger cost per person reached for advertisers than budget-splitting.
Bottom row of \autoref{fig:vrs_vs_split_by_both_imp_cpm} illustrates this,
  with VRS resulting in higher CPP than budget-splitting. %

Taken together,
  these results show that for mitigating bias in ad delivery simply splitting a campaign's budget across demographic groups would be both more effective in terms of people reached and in terms of cost than VRS.
\section{Discussion and Open Questions}
  \label{sec:future_work}

Our findings raise several questions that should guide
  future discussion on fairness interventions in ad delivery:
  considering alternatives to VRS's approach that provide better
  trade-offs between fairness, utility and transparency,
  giving external reviewers more capabilities to conduct effective audits,
  and giving users more control over what high-stakes ads they
  are shown.

\textbf{Are there better trade-offs between fairness, utility and transparency?}
The first is exploring approaches that provide better trade-offs
  between fairness and utility while also being transparent.
We do not propose budget-splitting (from \autoref{sec:method2})
  as a definitive solution, but it offers a useful alternative to Meta’s VRS implementation.
First, it shows that VRS’s higher advertiser costs and reduced user reach
  are not an inherent limitation of all fairness interventions.
Thus, similar to searches for better non-discrimination solutions that continue to serve business needs in other domains~\cite{Black2024, Laufer2025, finocchiaro2021bridging}, more efficient strategies than VRS for mitigating discrimination in ad delivery should be explored.
%
%
%

Second, it demonstrates that a much more transparent and interpretable fairness intervention, that performs no worse than VRS, is feasible. 
The advertisers themselves may not be able to split the budget according to demographic groups, for example, because they may lack access to the demographic composition of their audiences and because
  splitting a budget between the intersection of numerous demographic attributes
  can be a challenge. 
However, Meta could employ attribute inference in the same way they do for VRS, and thus realize this approach on advertisers' behalf. 
Moreover, Meta already offers advertisers numerous tools that leverage the platform's data to automatically configure advertisers' campaigns and split the budgets on their behalf: such as campaign budget optimization~\cite{campaignbudgetoptimization} and Advantage+~\cite{advantageplus}.
Thus, extending such efforts to fairness appears to be a path that does not require substantial engineering effort and would fit well into paradigms and tools advertisers are familiar with.
Finally, when the advertisers may have better information about their audiences than the advertising platform, the budget-splitting approach may do away with the need for platform's inference and result in more equitable outcomes~\cite{gelauff2020advertising}. 

We hope our experimental results based on the budget-splitting approach (\autoref{sec:experiments_method2}) are convincing in motivating the platforms to look for better, cheaper, and more transparent implementations for the desired fairness outcomes, and for regulators to push for them.

\textbf{What capabilities should external reviewers have?}
Our work also raises questions about what capabilities and level of access
  external reviewers should be given to provide sufficient
  oversight over platforms.
As discussed in \autoref{sec:verification},
  the current external reviewer mandated by the settlement only has access to
  aggregate ad delivery reports provided by Meta and does not have
  access to internal data and experimentation tools that would
  allow for independent verification of compliance.
The limited access raises concerns about whether reviewers
  can robustly detect noncompliance or evaluate the broader
  impacts of algorithmic interventions like VRS.
Future regulatory efforts need to consider these limitations and
  explore what additional levels of access are necessary for
  effective oversight~\cite{casper2024black}.
Potential approaches include giving auditors privacy-preserving
  access to internal data such as the output of algorithms
   platforms use to calculate the total value of ads used to determine
  ad auction winners~\cite{Imana2023},
  and providing means for auditors to perform socio-technical audits
  that study the impact of algorithmic interventions from the
  perspective of users~\cite{Lam2023}.

\textbf{What controls to offer users for high-stakes ads?}
Finally, beyond addressing limitations of VRS or introducing
  alternative fairness interventions,
  a more fundamental question is whether platforms should
  give users greater control over how they receive high-stakes ads.
Currently, platforms opaquely optimize ad delivery for ``relevance'' to users, with what is considered relevant chosen by the platform using imperfect signals of user actions, rather than by their true interests~\cite{kleinberg2024challenge}. 
One possible solution is to allow users to turn off
  relevance optimization for high-stakes ads in HEC domains,
  while continuing to use relevance optimization in the delivery
  of other ads such as entertainment and product ads.
There are growing efforts with goals for enacting values into recommender systems~\cite{stray2024building}, many of them focused on giving users greater control~\cite{fukuyama2020report, rajendra2022forgetful}, and some of the regulatory and policy effort may be well-spent on mandating approaches that improve user agency.

\section{Conclusion}
In this work, we evaluate the Meta-DoJ settlement and Meta's VRS implementation of it.
We identify critical gaps in the settlement requirements that
  allow for an implementation that does not improve access to opportunities
  for individuals.
We show that while VRS's implementation reduces variance as required by the settlement terms,
  it leads to fewer unique individuals being exposed to opportunity ads and
  increased costs for advertisers.
We demonstrate that alternative strategies,
   such as budget-splitting,
   can achieve better outcomes, illustrating the sub-optimality of Meta's chosen approach and offering clues as to the possibility of improvement.
 We propose potential areas for improvement in the settlement terms and VRS's effectiveness,
   such as incorporating reach-, rather than impression-, focused metrics,
   and having Meta share the cost of fairness intervention.
Our work contributes evidence for the need in increased transparency and independent evaluations of platform's efforts towards mitigating discrimination.%

\begin{acks}
This work was funded in part by the
   National Science Foundation grants CNS-1956435,
  CNS-2344925,
  and CNS-2319409, %
  and by the Alfred P. Sloan Research Fellowship for A. Korolova.
We thank our reviewers and Mitra Ebadolahi of Upturn for helpful feedback.
\end{acks}

\subsection*{Errata}

In~\autoref{sec:method1_exp_utiliery_user_adver} of the original version published at FAccT 2025,
  we miscalculated how VRS affects reach (\autoref{fig:per_group_impressions})
  and cost per 1000 reach  (\autoref{fig:vrs_change_in_cost}) for Male demographic attribute.
In this revision, we have fixed the figure and the associated text.
The change does not affect our conclusions.

\balance
\bibliographystyle{ACM-Reference-Format}
%
%
%

\newpage
\appendix

\section*{Appendix}
\section{Distribution of Impressions/Spend on Meta}
  \label{sec:meta_ad_dist}
\begin{figure}[b!]
\centering
    \centering
    \includegraphics[width=0.9\linewidth]{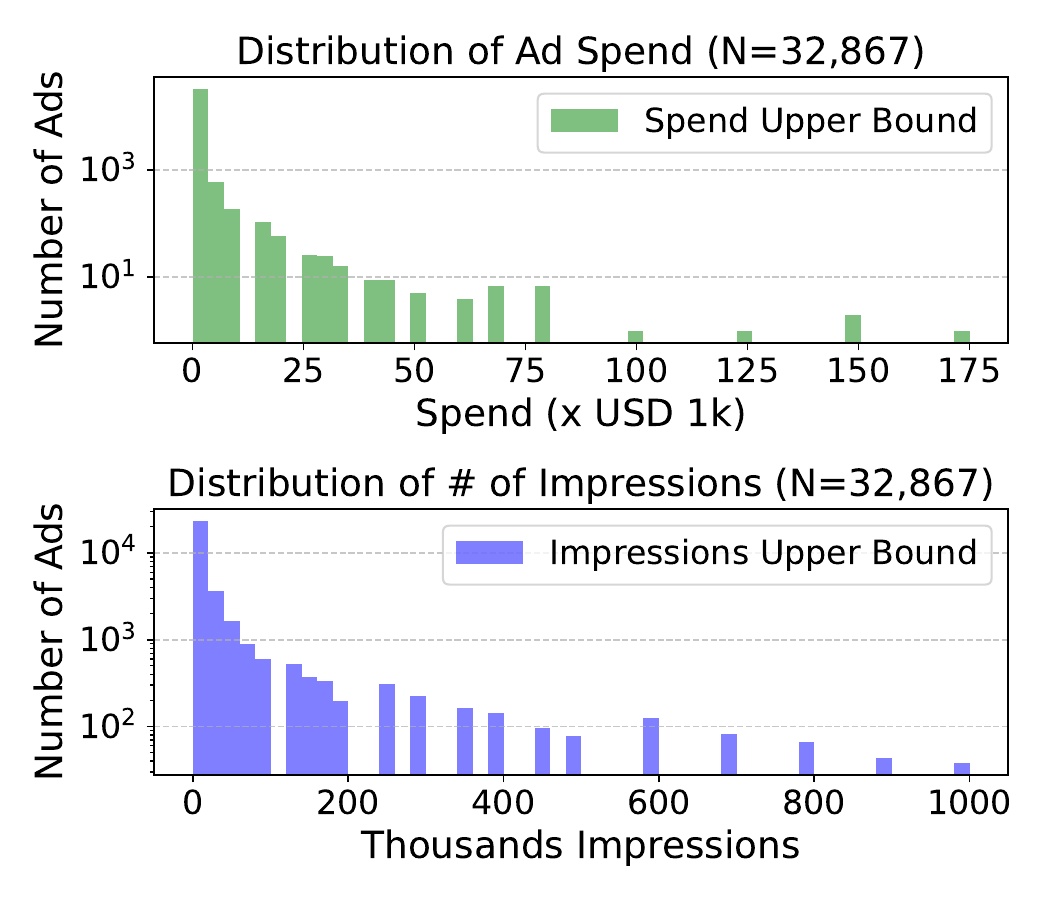}
 \caption{Distribution of ad spend and number of impressions for a sample of political ads from Meta's ad library.}
 \label{fig:powerlaw}
\end{figure}

In~\autoref{sec:coverage_limitation},
  we used data from Meta Ad Library to illustrate the impact of selective
  application of VRS to small advertisers for meeting the coverage
  requirements of the settlement.
\autoref{fig:powerlaw} shows the distributions of spend and impressions for the sample of ads we used
  for our analysis.
Both figures exhibit a heavy-tailed power-law distribution~\cite{Manning2008IR,powerlaw}:
  the majority of ads have low spend and receive relatively few impressions,
  whereas a long tail of high-budget ads 
  receive a very large number of impressions.
 
\section{Using Location as a Proxy for Race}
  \label{sec:loc_proxy}
  
In~\autoref{sec:ad_audience},
  we briefly mentioned that we build our audiences
  in a way that allows us to infer race from the location
  of ad recipients.
We next give more details on how we implement this proxy,
  following the approaches
  in prior external audits of ad delivery ~\cite{Ali2019a, Imana2021, Imana2024}.
  
Meta reports breakdown of ad recipients by gender,
  age and District Market Areas (DMAs), but not race.
Therefore, an auditor can build audiences
  using (race, DMA) pairs from NC voter data
  so that one can infer the racial breakdown of ad impressions based
  on the DMA breakdown reported by the platform.
The list of DMAs in NC split in to two parts, and include only Black
  individuals from one subset of DMAs and only White individuals
  from the other subset of DMAs.
Depending on which DMA an ad was shown,
  one can then infer whether it was shown to a White or Black person.
We use this same approach in our methodology to determine
  the racial breakdown of ad impressions.
To ensure location does not skew results, prior work replicates all
  experiments on “flipped” audiences, where the
  DMAs used to select groups are reversed~\cite{Ali2019a, Imana2024}.
We omit this step to avoid doubling the costs of our work,
  and because prior work found that flipping audiences produced similar
  results~\cite{Imana2024}.

\section{Privacy in VRS and its Implications}
  \label{sec:DP_vrs_overview}

In \autoref{sec:background}, we omitted details regarding
 steps that Meta's VRS implementation takes when measuring variance
 to protect the privacy of its users.
We now briefly describe these steps and their potential implications for VRS's ability to mitigate discrimination and for auditors' ability to verify Meta's compliance with established metrics.

Our key insights are that the privacy threats that necessitate taking the privacy-preserving steps are not clearly articulated by Meta, while the privacy-preserving steps complicate meaningful verification of compliance, even by the chosen third-party auditor.

\subsection{Differential Privacy in VRS}
%

%

\subsubsection{Use of Differential Privacy in VRS Implementation.}
  \label{sec:differential_privacy_in_VRS}
Meta aims to ensure differential privacy (DP)~\cite{dwork2006calibrating} both in the intermediate steps of the VRS module that adjusts the advertisers' bids up or down and in the compliance data it reports to the third-party auditor.
To do so it first adds carefully calibrated noise to the counts of impressions from each demographic group after the delivery of every $k$ impressions; and only these noisy counts are then used in the variance calculation and decisions for bid adjustments by the module~\cite{Timmaraju_2023, FacebookvsHUD2023TR}.
Furthermore, it adds calibrated noise to the actual impressions data it reports to the auditor.

In contrast to best-practices advocated by the DP community for differentially private deployments~\cite{dwork2019differential, tang2017privacy, wood2018differential}, neither the value of $k$ nor the target differential privacy guarantee parameter $\epsilon$ is shared in Meta's documentation describing VRS or with the auditor. 

\subsubsection{Implications of use of Differential Privacy on Fairness}

While DP provides a rigorous privacy protection,
  it is well-known that the noise it adds can impact fairness evaluations, and may introduce biases that are disproportionately distributed among demographic groups (see, e.g.~\cite{pujol2020fair, cummings2019compatibility, fioretto2025differentially}).
For example, prior work on DP in the U.S.~Census has
  demonstrated that the noise injection can
  disproportionately result in larger measurement error for smaller demographic groups~\cite{Asquith2022}.
It is not clear from VRS's documentation how Meta
  manages these potential trade-offs and what role the use of DP may play towards achieving the settlement goals.

\subsubsection{Implications of use of Differential Privacy on Auditability}\label{sec:auditability-app}
The application of DP at multiple points in the system,
  without a clearly articulated threat model,
  also reduces the ability of independent researchers to audit VRS.
Meta acknowledges that the DP approach increases the statistical variance
  of the system's measurements~\cite{Timmaraju_2023},
  allowing for a layer of plausible deniability where deviations observed by
  external researchers could be attributed to DP noise
  rather than underlying biases in ad delivery.
  
The DP noise also affects the ability of Guidehouse,
  the third-party reviewer mandated by the settlement~\autoref{sec:verification},
  to verify Meta’s reported numbers.
Even if Guidehouse
  was to conduct its own experiments,
  discrepancies between their findings and Meta's reported numbers could be
  attributed to noise from DP, making it challenging to verify the authenticity
  of the reported compliance metrics.
To overcome these challenges or incorrect variance calculations as a result of DP,
  we recommend future settlements require certification of application of DP, along the lines proposed by~\cite{bell2024certifying}. 

\subsection{BISG in VRS}
   \label{sec:bisg}
 As discussed in \autoref{sec:settlement},
   VRS measures variance by race using BISG,
   which outputs probabilistic estimates
   that a person belongs to a given racial group based on their surname and zip code~\cite{metatech, Elliott2009}.
Meta's implementation of BISG uses 50\%
  probability threshold to assign estimated race
  based on the group to which BISG assigns the highest probability~\cite{vrscomplianceJune24}.

The reliance on BISG may be problematic for a number of reasons, starting with accuracy.
The external reviewer in its audit report concludes that the choice of BISG threshold
  ``may have an impact on Variance and Coverage'',
  but that the 50\% threshold is reasonable because it is
  considered best practice~\cite{vrscomplianceJune24}.
Furthermore, existing evidence suggests that attribute inference tools such as BISG
  can be highly inaccurate~\cite{Ashurst2023, Lockhart2023},
  and that applying it to fairness in ad delivery systems without
  accounting for inference error
  may lead to underestimating disparities~\cite{Imana2025} or suboptimal outcomes~\cite{gelauff2020advertising}.
Therefore, VRS's design can benefit from incorporating
  noise-tolerant algorithms
   that directly account for uncertainty in demographic
  attributes~\cite{Mehrotra2021, Ghosh2021, Mehrotra2022, Ghosh2023}.

Finally, it appears that the inferences made by BISG, are the data whose privacy Meta aims to protect by using DP (see \autoref{sec:differential_privacy_in_VRS}), and thus its use directly implicates the difficulties for auditability discussed in \autoref{sec:auditability-app}.
Meta's privacy threat model for the VRS implementation is not clearly articulated; Meta's academic paper~\cite{Timmaraju_2023} only states they use DP for variance measurement to
  ``address various common issues such as privacy attacks discussed in \cite{Dwork2017}''. In our assessment, Meta does not sufficiently articulate how these issues may map to the VRS context, especially when the data is not shared with the advertisers, but only with the auditor under (presumably) very strict legal terms, including prohibition of reidentification. 
Another Meta technical report states their motivation for using DP with BISG is to ``prevent reidentification'' of
   individuals' estimated race attributes ~\cite{metatech}.
   It is unclear why this estimate needs reidentification protection, given that the BISG algorithm is public and thus inference of estimated race attributes for any given individual whose last name and zip code are known can be made by anyone.




\section{Understanding Skew in VRS's Delivery Ratio}
   \label{sec:skew_in_delivery_ratio}

\begin{figure}[b!]
  \centering
    \includegraphics[width=0.9\linewidth]{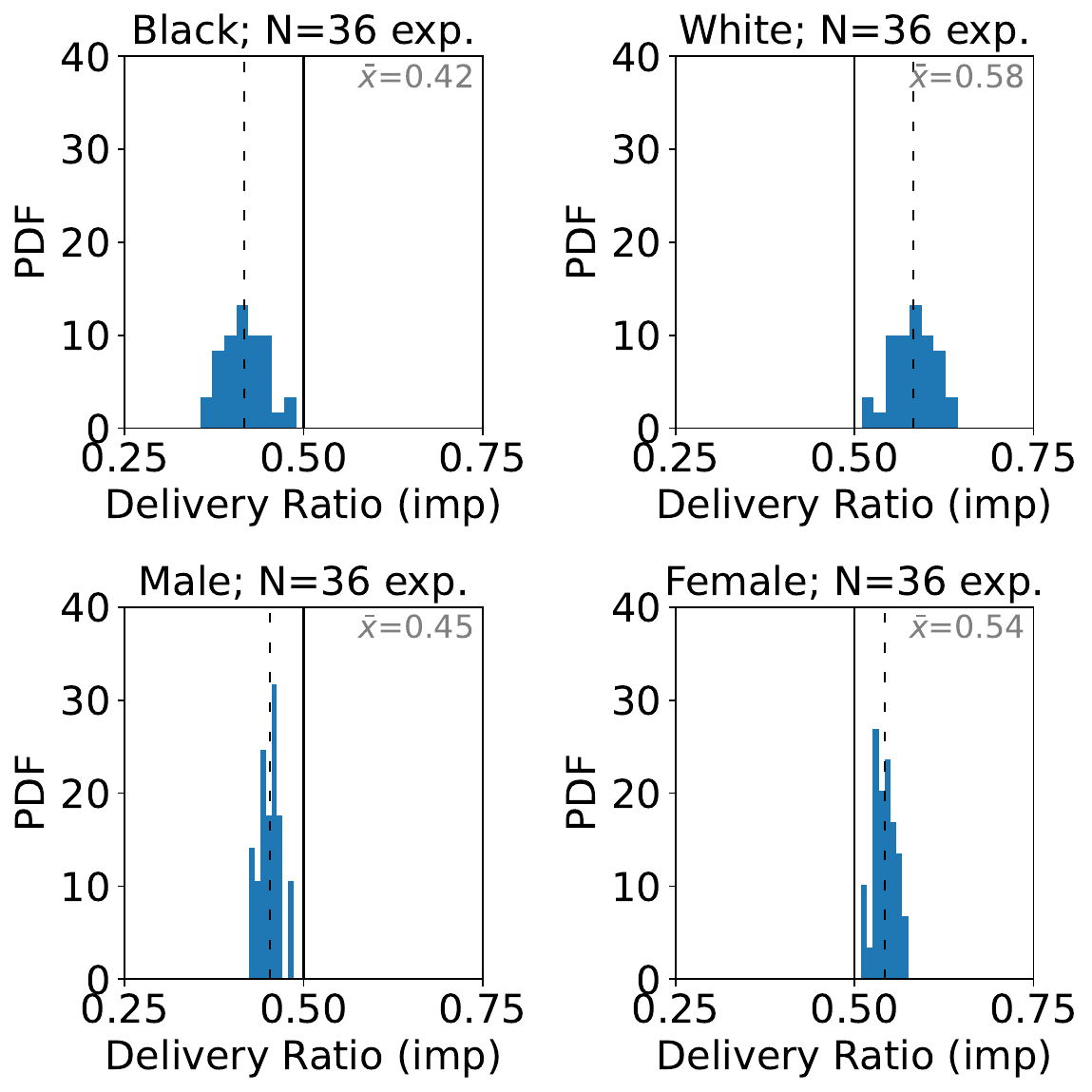}
    \caption{Delivery ratio by race and gender for all VRS-enabled ads. We use the mean delivery ratio for each demographic group (shown by $\bar{x}$ and vertical dotted line in the figures) 
    as an estimate for what eligible ratio VRS uses to reduce variance.}
     \label{fig:eligible_ratio_est} 
\end{figure}

In \autoref{sec:method1_exp_eligible}, we saw the delivery ratio for VRS
  does not match the even demographic split in the audiences we target.
In particular,
  our custom audiences contain equal number of individuals from all demographic groups (see \autoref{sec:ad_audience}),
  but we found the delivery ratios to be skewed by race (0.42 for Black; 0.58 for White) and  gender (0.45 for male; 0.55 for female)
  even after VRS's intervention.
\autoref{fig:eligible_ratio_est} illustrates this result,
  where the mean delivery ratio we use to estimate the eligible ratios is
  indicated by $\bar{x}$ and the vertical dotted lines.
  
We next show this gap can be explained by differences in the matching rates across
  demographic groups in our Custom Audiences.
For all audiences we used in our experiments in \autoref{sec:method1_exp_eligible},
  we first get the post-matching audience size for each group by uploading the audience list for each group separately.
Meta provides an API for querying how large a Custom Audience
  is after they match the information in the list we upload with real user accounts.
The API provides the sizes as a range for privacy reasons,
  so we take the midpoint of the range as an estimate.
We then use the estimated post-matching audience sizes to calculate the
  fraction of people included from each group.

The mean post-matching fractions we observe for each group
  are 0.43 for Black, 0.57 for White, 0.45 for male and 0.55 for Female,
  which closely match the delivery ratios we observed in \autoref{fig:eligible_ratio_est}.
Therefore, even though we upload an audience with an equal number of
  individuals from each group, the post-matching audience sizes can still be uneven.
This imbalance in matching rates skews the eligible ratio that VRS uses as a baseline,
  and explains the skewed delivery ratio we observe when VRS is enabled.

\end{document}